\documentclass[twocolumn,preprintnumbers,amssymb,amsmath,superscriptaddress,letterpaper]{revtex4}[10pt]

\usepackage{graphicx}
\usepackage{dcolumn}
\usepackage{bm}
\usepackage{natbib}
\usepackage{pstricks}
\usepackage{epsfig}

\newcommand{\be}{\begin{equation}}
\newcommand{\ee}{\end{equation}}
\newcommand{\bea}{\begin{eqnarray}}
\newcommand{\eea}{\end{eqnarray}}

\begin{document}

\title{Searching for the continuum spectrum photons correlated to the 130 GeV $\gamma$-ray line}

\author{Ilias Cholis}
\email{ilias.cholis@sissa.it}
\affiliation{SISSA, Via Bonomea, 265, 34136 Trieste, Italy}
\affiliation{INFN, Sezione di Trieste, Via Bonomea 265, 34136 Trieste, Italy}
\author{Maryam Tavakoli}
\email{tavakoli@sissa.it}
\affiliation{SISSA, Via Bonomea, 265, 34136 Trieste, Italy}
\affiliation{INFN, Sezione di Trieste, Via Bonomea 265, 34136 Trieste, Italy}
\author{Piero Ullio}
\email{ullio@sissa.it}
\affiliation{SISSA, Via Bonomea, 265, 34136 Trieste, Italy}
\affiliation{INFN, Sezione di Trieste, Via Bonomea 265, 34136 Trieste, Italy}

\date{\today}

\begin{abstract}
Indications for a $\gamma$-ray line(s) signal towards the Galactic center at an energy of about
130 GeV have been recently presented. While dark matter annihilations are a viable candidate for this signal,
it is generally expected that such a flux would be correlated to a $\gamma$-ray component with
continuum energy spectrum due to dark matter pair annihilating into other Standard Model particles.
We use the $\gamma$-ray data from the inner $10^{\circ} \times 10^{\circ}$ window to derive limits 
for a variety of DM annihilation final states.
Extending the window of observation, we discuss bounds on the morphological shape of a dark matter signal 
associated to the line,
applying both standard templates for the dark matter profile, such as an Einasto or a NFW profile, and
introducing a new more general parametrization. 
\end{abstract}

\maketitle

\section{Introduction}
\label{sec:intro}

Dark Matter (DM) accounts for approximately 85$\%$  of the matter density of the Universe. 
Among the many possible scenarios on its nature, Weakly Interacting Massive 
Particles (WIMPs) thermally produced in the early Universe are compelling 
candidates. WIMPs have a very rich phenomenology: they may be produced at 
colliders as the LHC, they may be detected by their direct interactions with baryonic 
matter or indirectly by singling out a component in cosmic rays and $\gamma$-rays due
to their pair annihilation into Standard Model (SM) particles. 

Among the possible signals of DM annihilation/decay, the search for a monochromatic 
$\gamma$-ray flux in multiple directions is one of the indirect detection methods with 
cleanest signature. As a prime target for such a signal, there is the galactic center 
(GC) direction, given that the annihilation/decay rate of DM in the halo peaks in that
region.

Recently \cite{Weniger:2012tx} has suggested the detection of a line at 
$129.8 \pm 2.4 ^{+7} _{13}$ GeV with a 3.3 $\sigma$ significance in a wide window towards 
the GC.  Additionally \cite{Su:2012ft} has revealed an excess of $\gamma$-rays concentrated 
around the GC that, if modeled as a single line, is at $127.0 \pm 2.0$ GeV at 5.0 $\sigma$
significance.
A pair of lines with energies of $110.8 \pm 4.4$ and $128.8 \pm 2.7$ GeV can also explain 
the excess of $\gamma$-rays with 5.4 $\sigma$ significance. Apart from the morphological 
differences these results agree on the energy of the line. Also both signals are in 
agreement with the constraints on line searches from the \textit{Fermi} Collaboration
\cite{Ackermann:2012qk}.
Other authors have also suggested the presence of one \cite{Tempel:2012ey} line at 130 GeV 
or multiple lines \cite{Buchmuller:2012rc, Boyarsky:2012ca} 
at $\simeq$ 110 and 130 GeV towards the galactic center. 
The morphology of the line signal from both \cite{Weniger:2012tx} and especially from 
\cite{Su:2012ft} leads us to concentrate on DM annihilation rather than decay.

Most WIMP models that can give a line(s) signal do so with a production rate which is loop-suppressed. 
Typically the monochromatic $\gamma$-ray yield comes together with a $\gamma$-ray yield with 
continuum spectrum due to the annihilation, at tree-level, into other two-body SM final states, which
in turn hadronize and/or decay into the stable species, $p$, $\bar{p}$ $e^{\pm}$, $\nu$s and 
$\gamma$-rays. Given that we have a preferred direction (inner few degrees of the GC) and a preferred mass 
range (correlated to the $\sim$130 GeV line), for the DM explanation of the line(s), we
can place specific limits on the continuous $\gamma$-ray component from DM annihilation 
(see for example \cite{Cirelli:2009dv, Buckley:2012ws}).
Alternative explanations for the line signal have also been suggested in 
\cite{Aharonian:2012cs, Profumo:2012tr, Boyarsky:2012ca}.  

In section~\ref{sec:LimitsDM} we describe our method of calculating limits on the
continuous component for a variety of SM annihilation products. 
Additionally, using the $\gamma$-ray data in the energy range of the line(s) and assuming
it is a DM signal we can study the morphology of events at that energy in the sky by 
extending our observation window. We show these results in section~\ref{sec:DMprofile}.  
In section~\ref{sec:WinoAxion} we implement our limits for a specific DM scenario 
and give our conclusions in section~\ref{sec:Conclusions}.
\section{Extracting limits on DM annihilation from the Inner $10^{\circ} \times 10^{\circ}$}
\label{sec:LimitsDM}

In \cite{Su:2012ft} the angular extension of the $\gamma$-ray line(s) signal at 110-130 GeV is 
described by a gaussian with FWHM of $3^{\circ}-4^{\circ}$.
If that signal comes from DM annihilation, the same region must be used to study 
the room for a $\gamma$-ray yield with continuum spectrum due to DM annihilation into other 
final states piling up at energies below 130 GeV.
For that purpose we use the region of the inner $10^{\circ} \times 10^{\circ}$ box
($\mid b \mid < 5^{\circ}$, $\mid l \mid < 5^{\circ}$).

In this analysis we have calculated spectra using 3 years of \textit{Fermi} LAT 
$\gamma$-ray data, taken between August 2008 and August 2011. 
Using the FermiTools PASS7 (v9r23p1) ``ULTRACLEAN''
class of data that ensures minimal CR contamination
\footnote{http://fermi.gsfc.nasa.gov/ssc/data/analysis/scitools/}.
We bin all the $\gamma$-ray events with energies of 200 MeV and up to 500 GeV in 30
logarithmically spaced energy bins \footnote{We check that changing the number of energy bins 
(between 25-35 bins) 
and their exact centers influences the limits on the continuous component only by $\simeq 1 \%$ and 
the estimation of the best fit value for the line(s) by up to $30 \%$ which is within the 
1$\sigma$ stated uncertainty of \cite{Su:2010qj} and \cite{Weniger:2012tx} for the luminosity 
of the line. Our choice of 30 bins is made to include almost all the line signal in one bin while 
having a large number of bins with good statistics for the continuous component.}. 
We separately calculate the exposures (and fluxes using HEALPix \cite{Gorski:2004by})
for front and back-converted events, before summing their contribution to the total flux
within the windows of interest.

To account for the contribution from known point and extended sources within the windows of
interest we have used the 2 yr published catalogue (see \cite{Collaboration:2011bm}
and references therein). Since we care to extract conservative limits on DM annihilation from 
$\gamma$-rays with energy below 130 GeV towards the inner $10^{\circ} \times 10^{\circ}$ 
we ignore the contribution of the \textit{Fermi} Bubbles \cite{Su:2010qj} $/$ \textit{Fermi} haze
\cite{Dobler:2009xz}, \cite{Dobler:2011mk}; 
given that there are significant 
uncertainties on the exact morphology of these structures at lower latitudes \cite{Dobler:2011mk}.

The $\gamma$-rays in the window of the inner $10^{\circ} \times 10^{\circ}$ 
originate from a combination of sources.
There are 29 detected point sources centered in that window 
\cite{Collaboration:2011bm}, 2 close by extended sources that contribute
minimally \cite{2010ApJ...718..348A, Ajello:2011pq}, as well as
the diffuse $\gamma$-rays from inelastic collisions
of CRs with the interstellar medium (ISM) gas, from bremsstrahlung radiation off
CR electrons and from up-scattering of low energy photons of the interstellar 
radiation field (ISRF) from high energy CR $e$ (inverse Compton scattering).
Additionally many unknown dim point sources are expected to be located within 
that window.
Finally the possible $\gamma$-ray contribution from DM annihilations in the halo 
is expected to peak towards the galactic center.

These $\gamma$-rays can be the direct product of DM annihilations as
in the case of the monochromatic yield from the 2$\gamma$, $Z \gamma$ or $h \gamma$
final states, possibly matching 
the lines detected by \cite{Weniger:2012tx, Su:2012ft}. Also virtual internal 
bremsstrahlung (VIB) and final state radiation (FSR) in DM annihilation can give a very hard 
spectrum that can be confused as a line over an otherwise featureless power law 
spectrum \cite{Bringmann:2012vr, Profumo:2012tr}. The decay of mesons (predominantly 
$\pi^{0}$s), produced in the decay or hadronization
processes of the products of DM annihilation, can also lead to a significant 
contribution to the gamma-ray spectrum. This component, while typically harder 
than the background $\gamma$-ray spectra, is significantly softer than the VIB/FSR and 
can not be confused as a DM line in $\gamma$-rays. 
These contributions probe directly the DM annihilation profile and will be referred to, as
prompt $\gamma$-rays. 
Additionally, inverse Compton and bremsstrahlung $\gamma$-rays from the leptonic
final products ($e^{\pm}$)  of DM annihilation will also contribute in that window
since both the ISRF energy density and the ISM gas density peak in the inner part 
of the Galaxy. 
   
As we suggested in the introduction, we will concentrate only on the DM
annihilation case since the line(s) morphology is so confined that it favors
profiles of cuspy annihilating DM halos (see also our discussion in section
~\ref{sec:DMprofile}).

To compute the diffuse $\gamma$-ray background we use the DRAGON package 
\cite{Evoli:2008dv, DRAGONweb} 
\cite{Cholis:2011un} with a new ISM gas model \cite{Maryam}
that ensures good agreement with $\gamma$-ray data in that window and overall 
\cite{MaryamEtAl}. We ignore in this work the contribution form the dark gas 
whose uncertainties are though significant in the inner $5^{\circ}$ in latitude
\cite{2005Sci...307.1292G, FermiLAT:2012aa} but that would only add to the diffuse 
$\gamma$-ray background resulting in less room for DM annihilation originated 
$\gamma$-rays.

We study five individual modes/channels of DM annihilation:
$\chi \chi \longrightarrow W^{+}W^{-}$, $\chi \chi \longrightarrow b\bar{b}$,
$\chi \chi \longrightarrow \tau^{+}\tau^{-}$, 
$\chi \chi \longrightarrow \mu^{+}\mu^{-}$ and 
$\chi \chi \longrightarrow e^{+}e^{-}$. Typically, DM models have sizable branching ratios 
into more than one of these channels. The exact limits in such models can be recovered by linearly 
combing the limits from the above channels. Annihilations to $Z$ gauge bosons give very 
similar $\gamma$-ray spectra to those of $W^{+}W^{-}$ bosons 
and annihilations to top quarks -not on shell in these cases- similar $\gamma$-ray spectra of annihilations 
to $b$ quarks. Thus the constraints
to those channels can be taken to be the same (within $\simeq 10\%$) to those of 
the $\chi \chi \longrightarrow W^{+}W^{-}$ ($\chi \chi \longrightarrow b\bar{b}$).

Following \cite{Su:2012ft} we assume that a line at energy of $127 \pm 2$ GeV 
has been detected. The morphology of the excess 
is described by a bi-gaussian with FWHM of 4 degrees in both $l$ and $b$. 
That line can come from $\chi \chi \longrightarrow 2\gamma$ or 
$\chi \chi \longrightarrow Z\gamma$ or $\chi \chi \longrightarrow h\gamma$.
In \cite{Weniger:2012tx} a single line at $129.8 \pm 2.4^{+7}_{-13}$ 
GeV has been suggested.
Additionally the case where there are 2 lines centered at $128.8 \pm 2.7$ and $110.8 \pm 4.4.$ GeV 
has been indicated by
\cite{Su:2012ft}. 
In that case the lines come from either the combination of $2\gamma \& Z \gamma$ lines
or from the $Z \gamma \& h \gamma$ lines. 

We study both the case of a single line centered at 127 GeV and the case of 2 lines 
centered at 129 and 111 GeV.
The choice of mass depends on the exact origin of the line(s).
For a single line from $\chi \chi \longrightarrow 2\gamma$ the mass range of 
$122 < m_{\chi} < 132$ GeV is studied. For a single line from 
$\chi \chi \longrightarrow Z\gamma$ we study $137 < m_{\chi} < 145$ GeV and from
$\chi \chi \longrightarrow h\gamma$ we study the $149 < m_{\chi} < 157$ GeV mass 
range.
For 2 lines originating from $2\gamma \& Z \gamma$ we study $127 < m_{\chi} < 130$ and for the
case of $Z \gamma \& h \gamma$ lines $138 < m_{\chi} < 143$ GeV (for a two line signal form DM 
annihilation see also \cite{Rajaraman:2012db}).
The relevant ratio in the luminosity of the two lines is taken to be 0.7/1 for the 111/129 GeV 
lines.
We allow for $4 \%$ uncertainty in the determination of energy of the line(s) which leads to the
ranges of masses referred above, which is about 2$\sigma$ of the declared uncertainties 
of \cite{Su:2012ft} and \cite{Weniger:2012tx} \footnote{In \cite{Su:2012ft} 16 logarithmically 
spaced energy bins were used to analyze the data between 80 and 200 GeV. That choice results 
in the bins being separated by a geometric factor of 1.06. Based on that we take
a $4 \%$ uncertainty in the energy which is more than half an energy bin in \cite{Su:2012ft}.}.

For every choice of annihilation channel to continuum $\gamma$-rays, annihilation channel(s) 
to line(s) and DM mass, we first find the best fit values from the $\gamma$-ray data 
within $10^{\circ} \times 10^{\circ}$ for both the cross-section of the main annihilation 
channel (giving the continuum $\gamma$-rays) and separately for the annihilation
cross-section to the line(s) (2 d.o.f.). 

\begin{figure}
\hspace{-0.5cm}
\includegraphics[width=3.10in,angle=0]{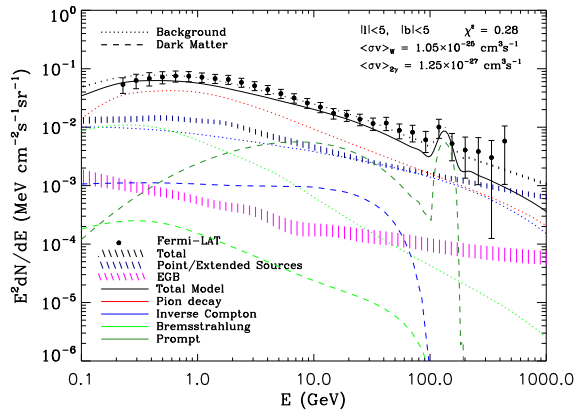}
\caption{Case of $m_{\chi} = 130$ GeV DM particle annihilating to a $W^{+}W^{-}$
pair with a cross-section of $1.05 \times 10^{-25}$ cm$^{3}$s$^{-1}$ and to a $2\gamma$
line with a cross-section of $1.25 \times 10^{-27}$ cm$^{3}$s$^{-1}$. We plot 
the $\mid b \mid < 5^{\circ}$, $\mid l \mid < 5^{\circ}$.}
\label{fig:GoodFit}
\end{figure}

In Fig.~\ref{fig:GoodFit}, we show a fit to the total $\gamma$-ray spectrum within 
our window of interest for the case of $\chi \chi \longrightarrow W^{+}W^{-}$,
with $m_{\chi} = 130$ and a single line coming from $\chi \chi \longrightarrow 2 \gamma$.
The fact that the best fit value for the cross-section is positive validates our claim 
of deriving conservative limits on DM annihilation, while the good agreement to the 
low ($E_{\gamma} < 1$ GeV) energies where the DM contributes at the few $\%$ level 
shows the good agreement of the physical model for the background to the low energy data.
    
\begin{figure*}[t]                                                                   
\includegraphics[width=3.75in,angle=0]{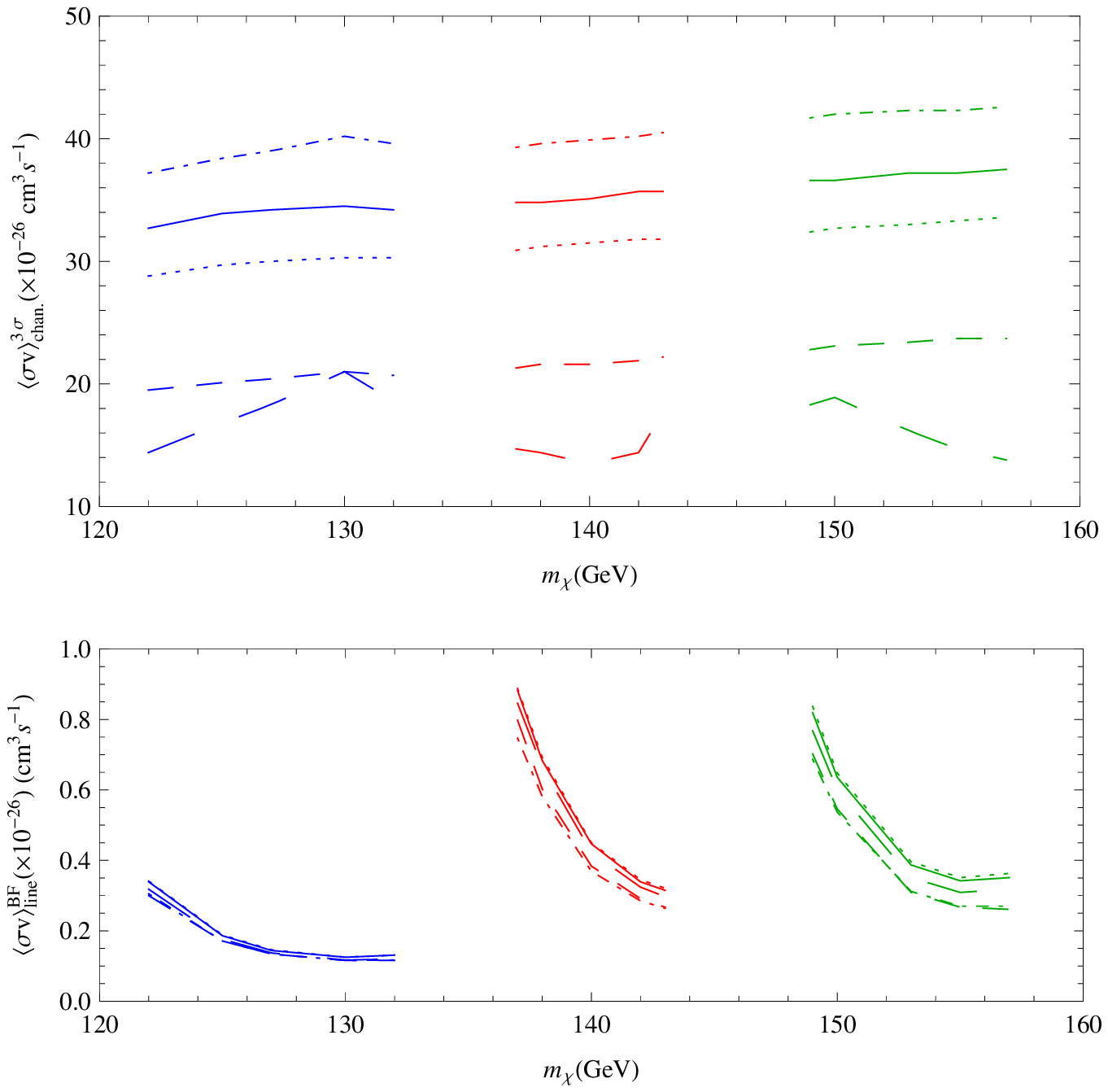}
\hspace{-1.5cm}
\includegraphics[width=3.75in,angle=0]{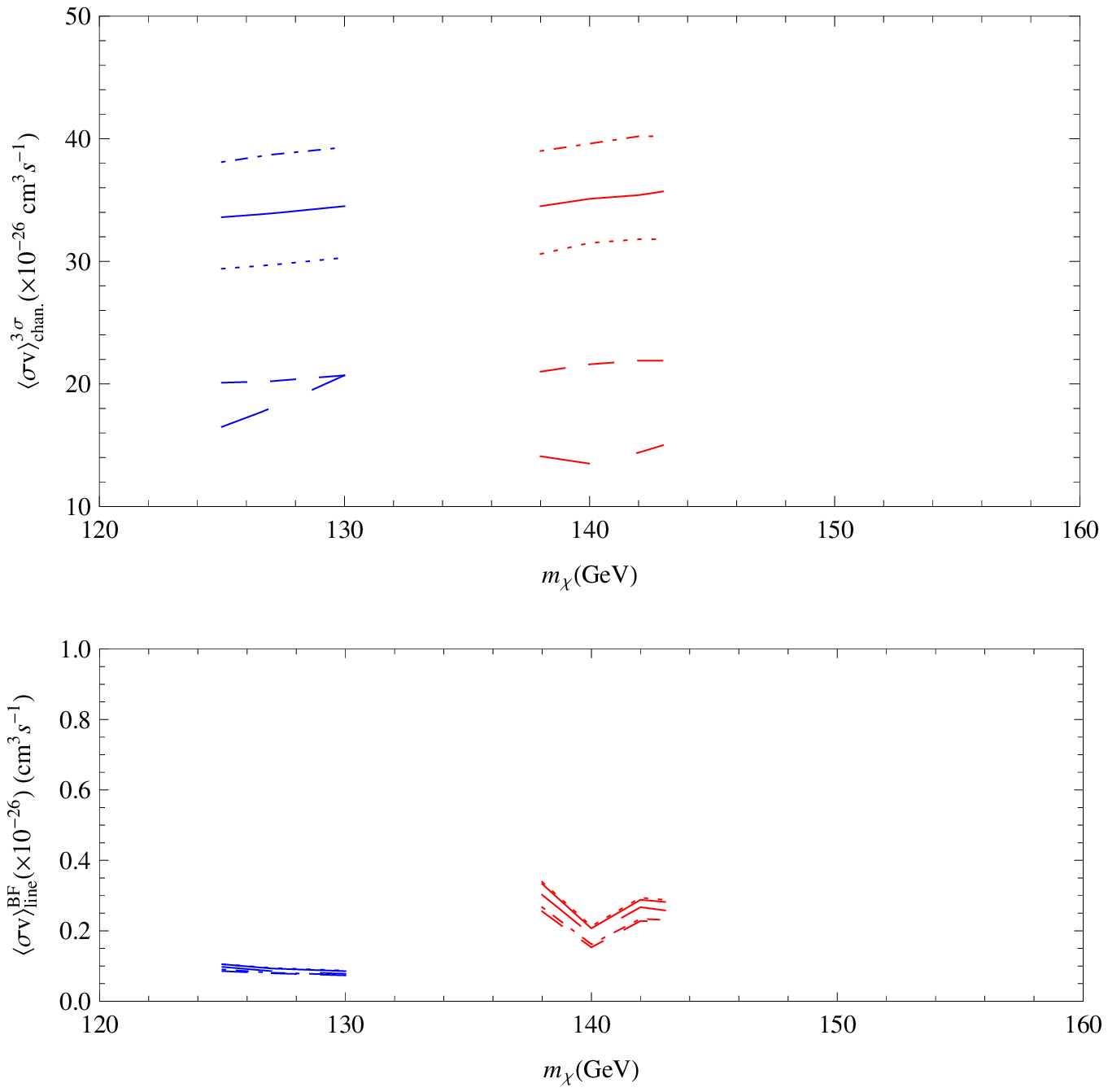}
\caption{Top: 3 $\sigma$ limits for annihilation into ``channel'', from region of 
$\mid b \mid < 5^{\circ}$, $\mid l \mid < 5^{\circ}$. 
We study five channels. $\chi \chi \longrightarrow W^{+}W^{-}$:
\textit{solid lines}, $\chi \chi \longrightarrow b\bar{b}$: \textit{dotted lines},
$\chi \chi \longrightarrow \tau^{+}\tau^{-}$: \textit{dashed lines},
$\chi \chi \longrightarrow \mu^{+}\mu^{-}$: \textit{dashed dotted lines} and
$\chi \chi \longrightarrow e^{+}e^{-}$: \textit{long dashed lines}.
Left: assuming a single line from $\chi \chi \longrightarrow 2\gamma$ (\textit{blue}), or
$\chi \chi \longrightarrow Z\gamma$ (\textit{red}) or $\chi \chi \longrightarrow h\gamma$
(\textit{green}). Right: assuming double lines from $\chi \chi \longrightarrow 2\gamma$ and
$\chi \chi \longrightarrow Z\gamma$ (\textit{blue}), from
$\chi \chi \longrightarrow Z\gamma$ and $\chi \chi \longrightarrow h\gamma$
(\textit{red}). Bottom: best fit values for the ahhinilation into line(s). 
For the double line case the annihilation best fit value refers to the cross-section 
for the highest energy line; the 111/129 GeV luminosity ratio is taken to be 0.7/1 . 
We use the Einasto DM profile of eq.~\ref{eq:Einasto} which gives $J/\Delta\Omega$ 
$= 1.21 \times 10^{24}$ GeV$^2$ cm$^{-5}$ (see text for more details).}
\label{fig:LimitsFull}
\end{figure*}

From the best fit value we then derive the 3$\sigma$ upper limits of the main annihilation
channel keeping the annihilation to the line fixed to its best fit value.
In Fig.~\ref{fig:LimitsFull} we show these 3$\sigma$ upper limits for the five 
annihilation channels to $W^{+}W^{-}$, $b\bar{b}$, $\tau^{+}\tau^{-}$,
$\mu^{+}\mu^{-}$ and $e^{+}e^{-}$.
We show limits for both the case of a single line Fig.~\ref{fig:LimitsFull} (top left)
and for a double line Fig.~\ref{fig:LimitsFull} (top right).
The exact choice of the origin of the line(s) and its energy(ies) has a subdominant 
effect on the limits in all channels except in the case of $e^{+}e^{-}$. That happens 
since the DM contribution of the main channel to the $\gamma$-ray spectrum is at energies 
bellow 100 GeV where the lines do not contribute.
The case of the $e^{+}e^{-}$ channel is an exception due to the very significant 
FSR component which peaks at $m_{\chi}$.
Thus the FSR component competes with the line(s) in the fit, making its limits sensitive 
to the exact assumptions on the line(s).    
We also give in Fig.~\ref{fig:LimitsFull} (bottom panels) the best fit values for the
line(s) for the relevant combinations of DM mass and channel.

The ISRF photon and gas densities have been fixed based on our background model.
The assumptions on these densities influence the inverse Compton and bremsstrahlung
components respectively. One can derive even more conservative limits on the DM
annihilation channels by considering only the prompt $\gamma$-ray contribution.

\begin{figure*}[t]
\includegraphics[width=3.75in,angle=0]{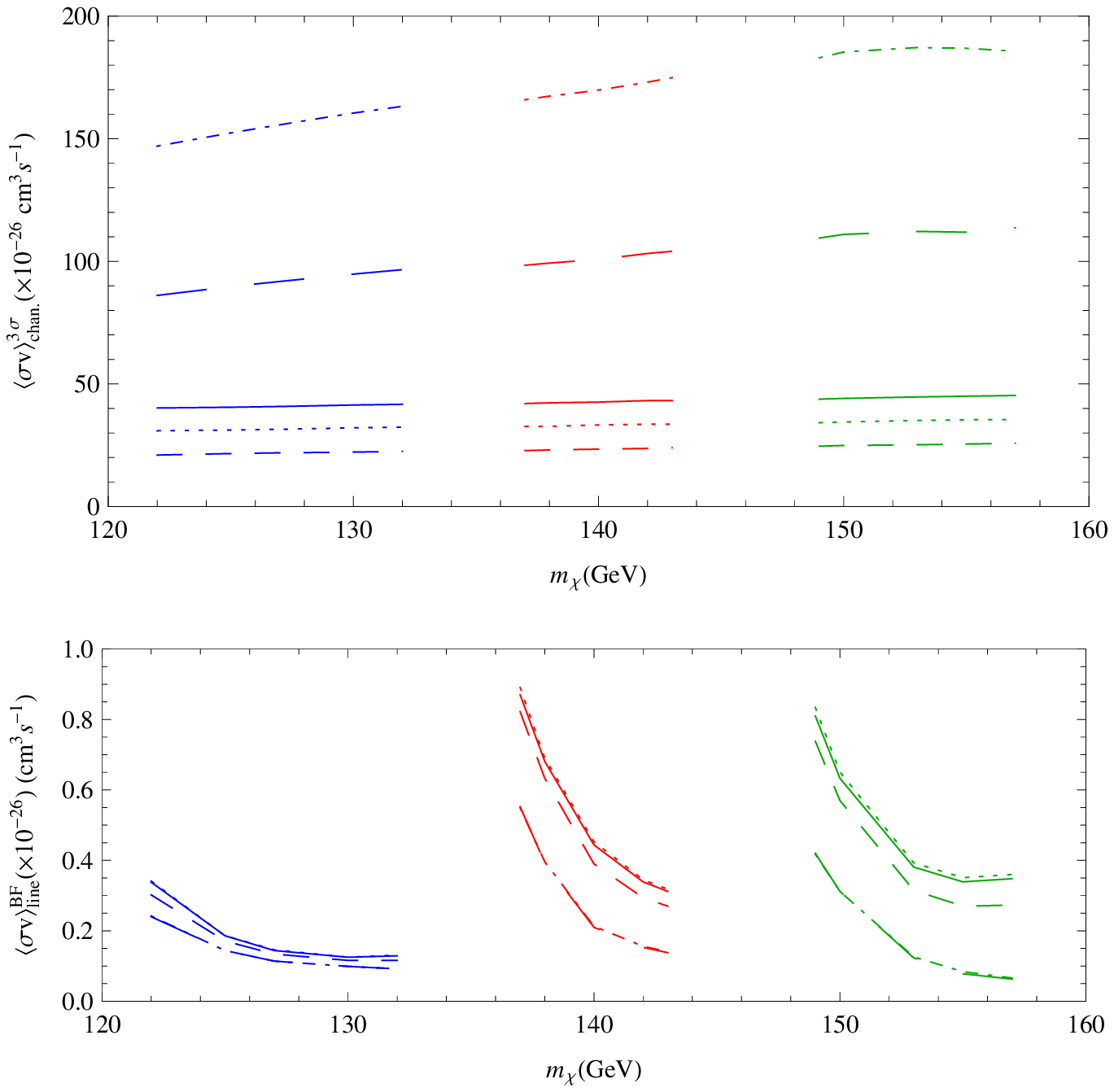}
\hspace{-1.5cm}
\includegraphics[width=3.75in,angle=0]{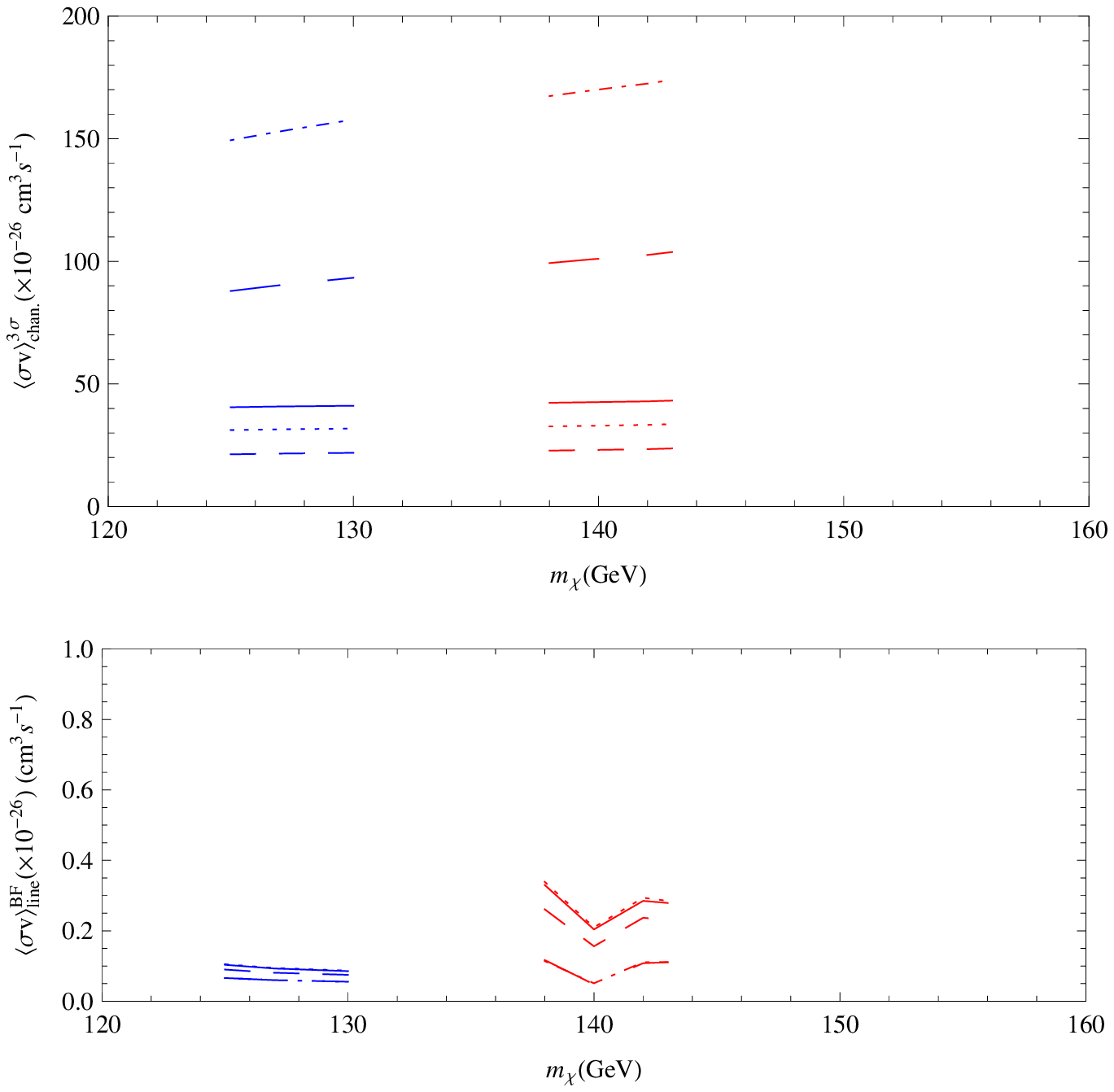}
\caption{Top: 3 $\sigma$ limits for annihilation into "channel" using only the prompt 
$\gamma$-rays from DM annihilation. We use the region of $\mid b \mid < 5^{\circ}$, 
$\mid l \mid < 5^{\circ}$. 
Left: assuming a single line. Right: assuming double lines.
Lines and colors as in Fig.~\ref{fig:LimitsFull}. 
Bottom: best fit values for the annihilation into line(s).
We use the Einasto DM profile of eq.~\ref{eq:Einasto}.}
\label{fig:LimitsPromptONLY}
\end{figure*}

In Fig.~\ref{fig:LimitsPromptONLY} we give the 3 $\sigma$ limits where only the prompt 
$\gamma$-rays from DM are taken into account. For the $W^{+}W^{-}$, $b\bar{b}$ and
$\tau^{+}\tau^{-}$ channels, for which the prompt $\gamma$-rays are the dominant component,
the limits become weaker only by $\simeq 10-20 \%$. For the $\mu^{+}\mu^{-}$, $e^{+}e^{-}$ 
modes on the contrary, since hard CR electrons are injected, their inverse Compton
and bremsstrahlung components are significant.
Thus if we ignore these diffuse components keeping only the prompt component, the 
3$\sigma$ limits become weaker by a factor of 4-5 in both channels.
  
The limits shown in Figs.~\ref{fig:LimitsFull} and~\ref{fig:LimitsPromptONLY} depend
on the DM profile assumptions.
We use here an Einasto DM profile:
\begin{equation}
\rho(r) = \rho_{Ein} \exp\left[-\frac{2}{R_{c}}*\left(\frac{r^{\alpha}}{R_{c}^{\alpha}}-1\right)  \right],
 \label{eq:Einasto}
\end{equation}
with $\alpha = 0.22$, $R_{c} = 15.7$ kpc and $\rho_{Ein}$ is set such that the local 
DM density is equal to 0.4 GeV
cm$^{-3}$ \cite{Catena:2009mf,Salucci:2010qr}.
That results in a J-factor from that window of $J/\Delta \Omega = 1.21 \times 10^{24}$ 
GeV$^2$cm$^{-5}$, where J factor is defined here as: 
\begin{equation}
J= \int_{\Delta \Omega} \int_{0}^{\infty} \rho_{DM}^{2}(s, \Omega) ds d \Omega,
\label{eq:Jfactor}
\end{equation}
with $s$ to be the distance along line of sight and $\Delta \Omega$ the angle
of observation.

A more cuspy DM profile would lead to stronger limits while a more cored (flat) in the
inner kpcs would lead to weaker limits. All the limits shown in Fig.~\ref{fig:LimitsPromptONLY}
and the limits for $W^{+}W^{-}$, $b\bar{b}$ and $\tau^{+}\tau^{-}$ in Fig.~\ref{fig:LimitsFull}
will change inverse proportionally (exactly or approximately) to the value of the $J$-factors
within that window, since the prompt component is dominant in these channels.
The same applies for the best fit values to the line(s). 
Thus these limits can be used for other DM profile assumptions once one properly takes into
account the different $J$-factor from that window. For the annihilation channels into 
$\mu^{+}\mu^{-}$ and $e^{+}e^{-}$ the limits in Fig.~\ref{fig:LimitsFull} have a dependence
on the DM profile that is more involved.

Finally since our aim in this paper is not to study the line itself but the accompanying
$\gamma$-ray fluxes for the DM case, we want to ensure that the exact line assumptions 
that we make do not influence our limits for the continuous component.
The 3$\sigma$ limits presented in Figs.\ref{fig:LimitsFull} and~\ref{fig:LimitsPromptONLY}
were derived with the cross-section to the line(s) to be the best fit value from the
fit to the $\gamma$-ray data within $\mid l \mid < 5^{\circ}$, $\mid b \mid < 5^{\circ}$.
Alternatively, we calculate the 3$\sigma$ limits for the same channels using for the cross-section 
to the line(s) such a value that gives the luminosity stated for the $4^{\circ}$ FWHM cusp of
\cite{Su:2012ft}, that is $(1.7 \pm 0.4)\times 10^{36}$ph/s or $(3.2 \pm 0.6)\times 10^{35}$erg/s. 
The difference in the values of the cross-sections 
to the line(s) between the two methods is $\simeq 30\%$ (at the same level with the stated 
uncertainty of \cite{Su:2012ft}).

In Table~\ref{tab:LineNorm} we present our limits \textit{on the continuous components} 
for these two alternative methods of evaluating the cross-section to the line(s) before 
deriving the limits. For the case where the cross-section value to the line(s) comes 
from the $\mid l \mid < 5^{\circ}$, $\mid b \mid < 5^{\circ}$ region fit, (denoted as 
\textit{``free''}) and for the case where the cross-section comes form the luminosity 
stated by \cite{Su:2012ft}.
We show all five channels for three masses characteristic for the three DM mass ranges
valid in the case of a single line at 127 GeV.     
\begin{table}[t]
\begin{tabular}{|c|c||c|c|c|}
\hline
Chan.&Line& $127$ GeV ($2\gamma$)& $140$ GeV ($Z\gamma$)& $150$ GeV ($h\gamma$) \\
\hline \hline
$W^{+}W^{-}$&Free& 34.2(40.8) & 35.1(42.6) & 36.6(44.1) \\
\hline
$W^{+}W^{-}$&Fixed& 34.5(41.4) & 35.4(43.2) & 37.2(44.7) \\
\hline
$b\bar{b}$&Free& 30.0(31.5) & 31.5(33.3) & 32.7(34.5) \\
\hline
$b\bar{b}$&Fixed& 30.3(31.8) & 31.8(33.6) & 33.0(34.8) \\
\hline
$\tau^{+}\tau^{-}$&Free& 20.4(21.9) & 21.6(23.4) & 24.1(24.9) \\
\hline
$\tau^{+}\tau^{-}$&Fixed& 20.7(21.9) & 21.9(23.7) & 23.4(25.2) \\
\hline
$\mu^{+}\mu^{-}$&Free& 39.0(155.7) & 39.9(169.8) & 42.0(185.4) \\
\hline
$\mu^{+}\mu^{-}$&Fixed& 41.1(156.3) & 40.2(167.7) & 42.3(184.5) \\
\hline
$e^{+}e^{-}$&Free& 18.3(91.8) & 13.5(100.8) & 18.9(111.0) \\
\hline
$e^{+}e^{-}$&Fixed& 18.3(92.1) & 13.5(99.3) & 19.2(110.4) \\
\hline \hline
\end{tabular}
\caption{3$\sigma$ upper limits on DM annihilation $\langle \sigma v \rangle \times BR$ 
to \textit{channel} (i.e the continuum part) 
in units of $\times 10^{-26}$ cm$^{3}$s$^{-1}$ using full (in parenthesis:only prompt) DM $\gamma$-ray
spectra within $\mid l \mid < 5^{\circ}$, $\mid b \mid < 5^{\circ}$. The line signal is 
taken to be either from its best fit value (\textit{free}) or \textit{fixed} using the luminosity
of \cite{Su:2012ft} (see text for more details).  The $J$-factor$/\Delta \Omega$ from this window 
is $1.21 \times 10^{24}$ GeV$^{2}$cm$^{-5}$.}
\label{tab:LineNorm}
\end{table}

The difference in the limits for all five channels and all masses between the two methods is 
at the $\simeq 1 \%$ level. The same results apply for the case of 2 lines (111 and 129 GeV).
Thus the exact luminosity assumptions for the line(s) can not influence our results on the 
continuous component.
 
We also find that changing our window of observation from $(\mid l \mid, \mid b \mid) < 5^{\circ}$, 
to $(\mid l \mid, \mid b \mid) < 3^{\circ}$, $4^{\circ}$ or $8^{\circ}$ our limits for the continuous
component (for the best fit value for the line(s)) change by up to $10 \%$ ($20\%$), with the limits 
from $(\mid l \mid, \mid b \mid) < 5^{\circ}$ being the strongest (see also work of 
\cite{Cohen:2012me, Buchmuller:2012rc}).

\section{Dark Matter Annihilation Signal Profile}
\label{sec:DMprofile}

Ref. \cite{Su:2012ft}, suggests that the line(s) signal can be morphologically fitted by a 
$4^{\circ}$ FWHM gaussian distribution or $3^{\circ}$ FWHM when using just the events 
and avoid making diffuse maps or masking out any part of the GC.  The author of \cite{Weniger:2012tx}
has suggested instead a wider region of best significance for the 130 GeV line.

Using ULTRACLEAN data class we address as well the matter of DM profile morphology
under the assumption that the line signal is of DM origin and that an associated 
continuous spectrum exists. 
   
Calculating the $\gamma$-ray spectral data within a wider region of the sky we can 
derive limits on the allowed annihilation cross-section for a specific assumption 
on the DM halo profile or vice versa for the DM halo profile properties for 
specific assumptions on the annihilation cross-section. 

Motivated by the $\simeq 130$ GeV energy of the $\gamma$-ray line, we consider a DM mass 
of $m_{\chi} = 130$ GeV annihilating to $W^{+}W^{-}$, with a cross-section to $2\gamma$ 
for the line.

\begin{figure*}[t]
\includegraphics[width=3.20in,angle=0]{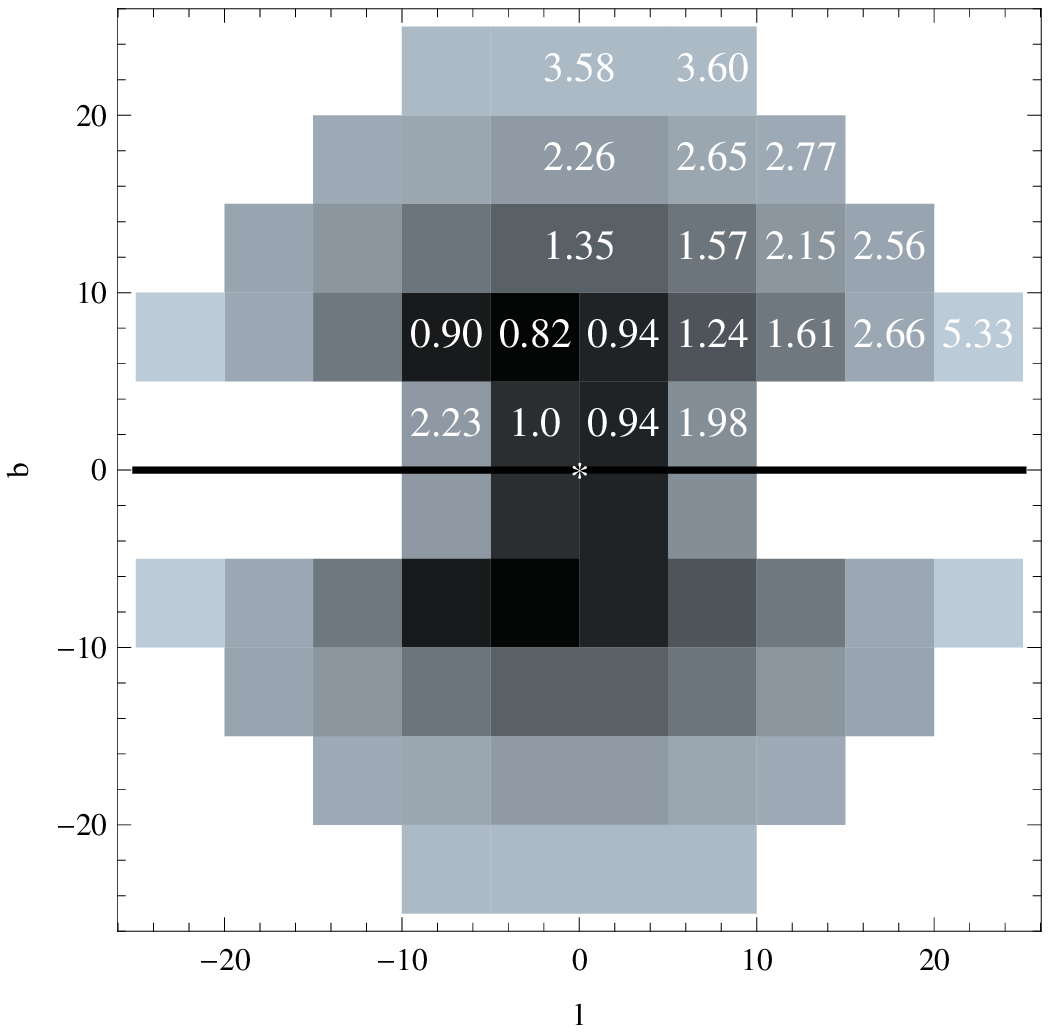}
\hspace{0.1cm}
\includegraphics[width=3.20in,angle=0]{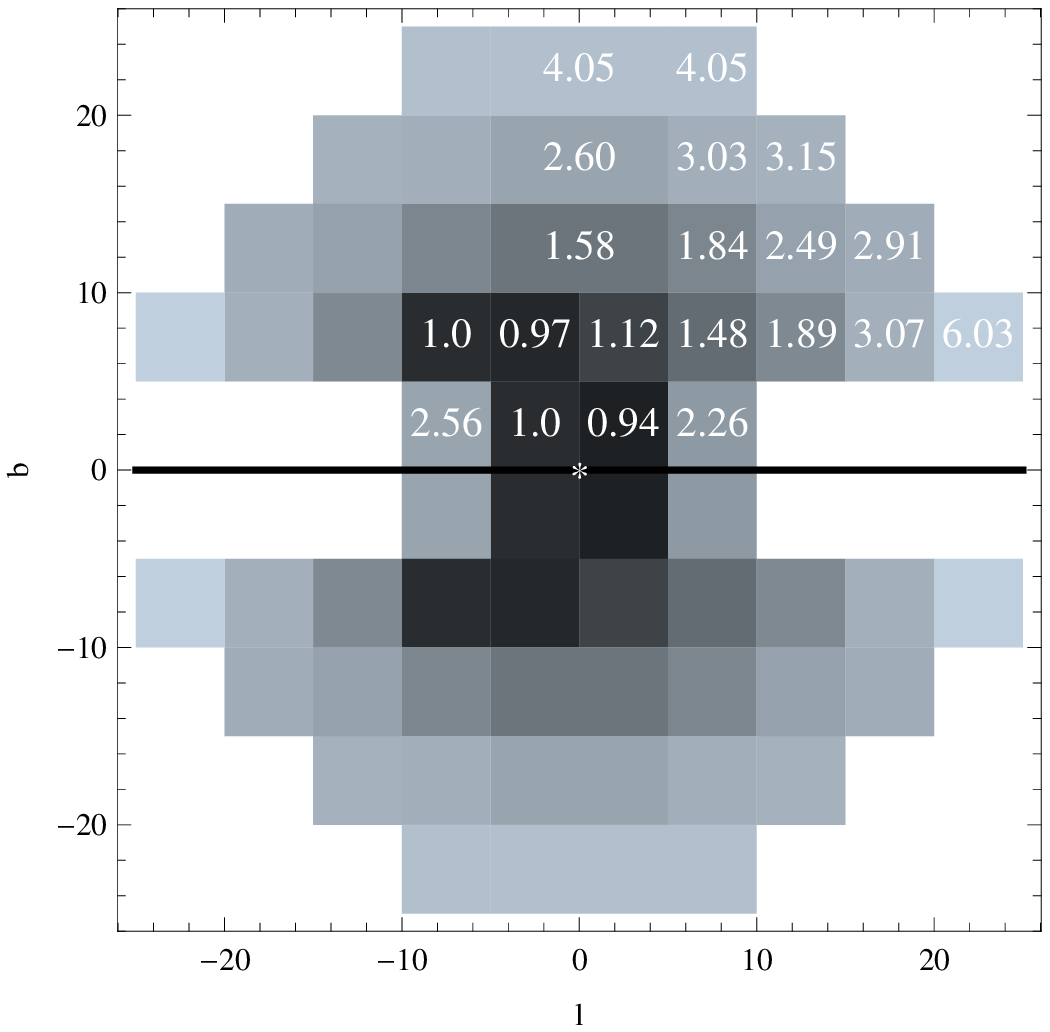}
\caption{Strength of limits on the annihilation cross-section to $W^{+}W^{-}$.
Left: Einasto profile. Right: NFW profile. Values are normalized to the 3 $\sigma$
limit on the annihilation cross-section of $\chi \chi \longrightarrow W^{+}W^{-}$
from $-5^{\circ} < l < 0^{\circ}$, $\mid b \mid < 5^{\circ}$ 
($\langle \sigma v \rangle^{3\sigma}_{-5^{\circ} < l < 0^{\circ}, \, \mid b \mid < 5^{\circ}}$
$=3.72 \times 10^{-25}$cm$^{3}$s$^{-1}$). Values smaller 
than 1 indicate stronger(lower) 3 $\sigma$ limit by the given (multiplication) factor.}
\label{fig:DMprofile}
\end{figure*}

We calculate the $\gamma$-ray spectra in the same energy binning as for the 
$10^{\circ} \times 10^{\circ}$ box described in section~\ref{sec:LimitsDM}.
We concentrate in the $\mid b \mid < 25^{\circ}$, $\mid l \mid < 25^{\circ}$
region where the annihilation from the halo is dominant. 
We break that region in 20 smaller windows, 8 of which in the region $\mid b \mid < 10^{\circ}$, 
$\mid l \mid < 10^{\circ}$ of $5^{\circ}$ size in $l$ and $10^{\circ}$ size in $b$. For 
$\mid b \mid > 10^{\circ}$, $\mid l \mid > 10^{\circ}$,  12 windows symmetrically placed 
with respect to $b=0^{\circ}$, $l=0^{\circ}$ each composed of 4 boxes $5^{\circ} \times 5^{\circ}$ size. 
These windows are also shown in Fig.~\ref{fig:DMprofile}.

In Fig.~\ref{fig:DMprofile}, we assume the DM profile to be either an Einasto 
as described in eq.~\ref{eq:Einasto}, (Fig.~\ref{fig:DMprofile}, left panel) or an NFW
profile (Fig.~\ref{fig:DMprofile}, right panel):
\begin{equation}
\rho(r) = \rho_{NFW} \left(\frac{R_{c}'}{r}\right) \left(\frac{1}{1+\frac{r}{R_{c}'}}\right)^{2},
 \label{eq:NFW}
\end{equation}
with $R_{c}' = 14.8$ kpc, $\rho_{NFW} = 0.569$ GeV cm$^{-3}$.
For both cases we fit the DM density to the locally measured value of 0.4 GeV cm$^{-3}$ 
\cite{Catena:2009mf}.
That results in a specific J-factor for each window.

We calculate as in section~\ref{sec:LimitsDM} the 3$\sigma$ limits on 
$\chi \chi \longrightarrow W^{+}W^{-}$ from each angular window. 
These limits include $\gamma$-ray background contribution that as in the 
$10^{\circ} \times 10^{\circ}$ box fits the lower energies. 

In Fig.~\ref{fig:DMprofile} we give the ratios of the 3$\sigma$ limit on each window
devided by the 3$\sigma$ limit from the window of $-5^{\circ} < l < 0^{\circ}$, 
$\mid b \mid < 5^{\circ}$:  
\begin{equation}
\frac{\langle \sigma v \rangle^{3\sigma}_{window}}{\langle \sigma v \rangle^{3\sigma}_{-5^{\circ} < l < 0^{\circ}, \, \mid b \mid < 5^{\circ}}}.
\label{eq:Ratio}
\end{equation} 
     
Values of the ratio in eq.~\ref{eq:Ratio} smaller than 1 indicate stronger 3$\sigma$ 
limits than that in the window of reference. 
We choose the window of $-5^{\circ} < l < 0^{\circ}$,
$\mid b \mid < 5^{\circ}$ as reference since -as also thoroughly discussed in 
\cite{Su:2012ft}- the majority of photons with energy $\sim 130$ GeV
come from that part. We find 38 photons with energy $104.5-135.7$ GeV 
(the one energy bin in our analysis that includes the line(s)) to come from the 
$-5^{\circ} < l < 0^{\circ}$, $\mid b \mid < 5^{\circ}$ window, while only 21
photons in the same energy bin come from the symmetrical to the GC,  
$0^{\circ} < l < 5^{\circ}$, $\mid b \mid < 5^{\circ}$ window.

In our simulations the DM halo profile is centered at $l=0^{\circ}$, $b=0^{\circ}$.
We find that our 3 $\sigma$ limits are stronger from the $0^{\circ} < l < 5^{\circ}$, 
$\mid b \mid < 5^{\circ}$ window than from the $-5^{\circ} < l < 0^{\circ}$, 
$\mid b \mid < 5^{\circ}$ one. We remind that these limits come from the continuous 
component (see discussion in section~\ref{sec:LimitsDM}), thus there is an indication
even from the continuous DM component that there is an excess of $\gamma$-ray fluxes 
from the $l<0$ side. That is seen with both the Einasto and NFW profiles.

We do not claim that such an excess is due to DM annihilations since it could equally 
correspond to an underestimation of the background in that energy range.
Though that can be the case also for the line \cite{Boyarsky:2012ca, Aharonian:2012cs}.
Yet, if the line is of DM origin and is off-center as suggested by \cite{Su:2012ft},
the strength of the derived limits may indicate an analogous effect in the continuum term. 

Once we move to $\mid l \mid > 5^{\circ}$ and $\mid b \mid < 5^{\circ}$ the strength of the 
3$\sigma$ limits drops. For the Einasto profile the limits become stronger at 
$-10^{\circ} < l < 5^{\circ}$, $5^{\circ} < \mid b \mid < 10^{\circ}$ and weaker in
all other windows. For the more centrally peaked NFW profile the only region where
slightly stronger limits are recovered is that of $-5^{\circ} < l < 0^{\circ}$, 
$5^{\circ} < \mid b \mid < 10^{\circ}$.  

Turning the perspective around, we discuss now what information can be extracted on the dark 
matter profile, having fixed the annihilation cross section to a reference value. Rather than introducing for the dark matter
density some functional form specified in terms of few parameters, as most usually 
done in the literature, we consider here a much more general approach. Our generic 
spherically symmetric profile is set via: {\sl i)} Specifying its value $\rho_i$ at the 
seven Galactocentric distances $r_i=R_0 \, \sin(\alpha_i)$, being $R_0$ the Sun Galactocentric 
distance and with $\alpha_i= 5^{\circ}, 10^{\circ}, 15^{\circ}, 20^{\circ}, 25^{\circ}, 45^{\circ} \& \,65^{\circ}$, 
namely at the spherical shell corresponding to the angular windows already introduced above, 
plus three higher latitude patches; {\sl ii)} Fixing the dark matter density at the local 
Galactocentric distance $r_8 \equiv R_0$ to $\rho_8 \equiv 0.4 \,{\rm GeV} {\rm cm}^{-3}$ 
and implementing a linear interpolation in a double logarithmic scale to retrieve the density 
profile between any two of these radii,  i.e. allowing for an arbitrary power-law scaling between 
any two $r_i$, with the only extra assumption of imposing that the profile is monotonically 
increasing for decreasing radius;  {\sl iii)} Assuming that the profile follows our reference 
Einasto model for $r>R_0$, a choice that has no impact on the analysis that follows. 

We refer again to the sample dark matter case introduced in Fig.~\ref{fig:GoodFit}, with $m_{\chi} = 130$~GeV and 
annihilating to $W^{+}W^{-}$. While the ratio between the cross section into $2\gamma$ to the one 
into $W^{+}W^{-}$ is fixed by the best fit value (in the specific case we found it to be about 0.012), 
the absolute value of the cross section scales with the inverse of the line of sight integration 
factor $J$ in the angular window $\mid b \mid < 5^{\circ}$, $\mid l \mid < 5^{\circ}$, in turn depending 
on the density profile within all the angular shells introduced above. 
After choosing a reference value for $\langle \sigma v \rangle$, we wish to derive how large a 
contribution to $J$ may come from each of the shells in our model, without violating the constraints 
coming from data in other angular windows. This gives an indication on how centrally concentrated the dark 
matter profile should be to provide a signal in the GC direction and, at the same time, to be consistent 
with data away from it. For this purpose we introduce the factors $J_i$ which are analogous to the 
$J$-factor introduced in Eq.~\ref{eq:Jfactor}, except for imposing that the density profile is constant 
below the radius $r_i$, namely $\rho(r<r_i) = \rho_i$. 

The analysis is performed scanning the parameter space defined by the values $\rho_i$ 
(with $i\in[1,7]$). For each model we compute line of sight integration factors corresponding to the angular 
regions displayed in Fig.~\ref{fig:DMprofile}, as well as for the regions at 
$25^{\circ} < \mid b \mid < 45^{\circ}$, $45^{\circ} < \mid b \mid < 65^{\circ}$ 
and $65^{\circ} < \mid b \mid < 85^{\circ}$, and $0^{\circ} \mid l \mid < 20^{\circ}$ 
(for all three latitude intervals). For all regions we can compare against the 
$3\sigma$ upper bound on the flux due to a dark matter candidate with mass 130 GeV and annihilating 
to $W^{+}W^{-}$ (conservatively including prompt emission only, while ignoring the 
radiative emission from the associated lepton yields). As upper bound to the monochromatic signal 
we consider instead the mean flux integrated in the energy bin [104.5,~135.7]~GeV, obtained 
again using the ULTRACLEAN data sample, and under the very conservative hypothesis of zero background 
from diffuse emission. Among models passing constraints, we search for configurations giving 
the maximum for the individual terms $J_i$ (we also implement the additional limit $J_1 \le J$, given 
that any additional contribution to the line of sight integral at radii $r<r_1$ is always neglected in our setup). 
Although we are dealing with a very large parameter space, finding the upper bounds to $J_i$, which we 
label $J_i^{max}$, is not exceeding expensive since one can show that, for each radial shell, they mostly 
correspond to the models with largest changes in profile slope between neighboring shells. 

\begin{figure}
\hspace{-0.5cm}
\includegraphics[width=3.10in,angle=0]{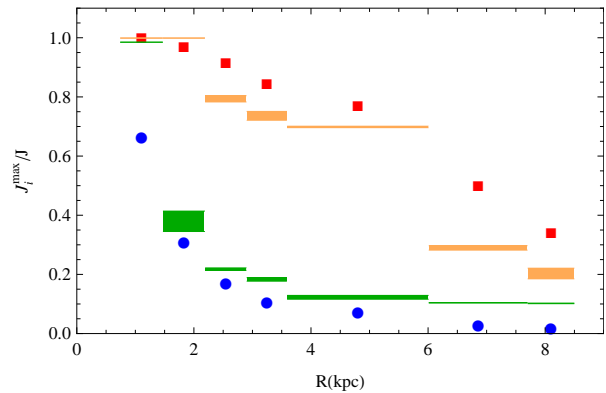}
\caption{Upper bounds on the partial $J_i$ factor, normalized to the total GC line of sight integration factor $J$,
for the parametric dark matter density profile introduced in the text.
The $J_i$ factor are computed assuming a constant dark matter density within the corresponding
radial shell $r_i$. The $J_i^{max}$ values displays are derived
implementing separately the limits on the monochromatic flux (lower green bands) and the continuum
spectrum (higher orange bands) derived from the other angular windows considered in the analysis.
Limits are shown as narrow bands since they refer to three different values of the annihilation cross
section, see the text for details.
Also shown are the values for $J_i/J$ for our reference Einasto profile (blue dots) and for a
cored Burkert profile (red squares); as it can be seen the Einasto profile is allowed, while the Burkert
shape is excluded.}
\label{fig:Jlimits}
\end{figure}

Ratios between $J_i^{max}$ and $J$ are shown in Fig.~\ref{fig:Jlimits}, where we display separately the $J_i^{max}$ 
found when applying the limit on the monochromatic flux and when implementing that from the 
component with continuum spectrum; limits are shown as (very narrow) bands since they were derived 
for three different values for $\langle \sigma v \rangle$: the ``thermal'' value  $3\times 10^{-26}$ cm$^{3}$~s$^{-1}$, 
the best fit value in case of our reference Einasto profile $1.05 \times 10^{-25}$ cm$^{3}$~s$^{-1}$, and ten times the thermal value
(this shows that dependence of our analysis on $\langle \sigma v \rangle$ is really very mild). 
For comparison, we plot also values of $J_i/J$ for our reference Einasto profile and for a Burkert profile, 
namely $\rho \propto 1/(r+R_c)/(r^2+R_c^2)$, with local dark matter density
$0.4$ GeV cm$^{-3}$ and core radius $R_c = 10$~kpc~\cite{Catena:2009mf}. As one can see the Burkert 
profile is excluded from both line and continuum components, while the line limits are giving stronger evidence towards 
the need for a more centrally concentrated dark matter profile. This is most probably related to the fact that the 
limits are derived in part from regions of the sky where the \textit{Fermi} Bubbles/haze, 
has been claimed to be needed; we do not try to include such component in our background model 
and most probably this translates into an extra room (or a less severe constraint) on the 
continuum emission from dark matter annihilations. On the other hand, the \textit{Fermi} Bubbles/haze are expected 
to play a marginal role at high energy, hence the sharper constraint from the line emissivity. 

\section{A specific example}
\label{sec:WinoAxion}
    
The limits that we show in Fig.~\ref{fig:LimitsFull} and~\ref{fig:LimitsPromptONLY} 
can be used (linearly combined) for a wide class of models. 
As a specific example we use our code to evaluate the constraints on a relevant model
presented for the explanation of a 130 GeV line \cite{Acharya:2012dz}.

\begin{figure}
\hspace{-0.5cm}
\includegraphics[width=3.10in,angle=0]{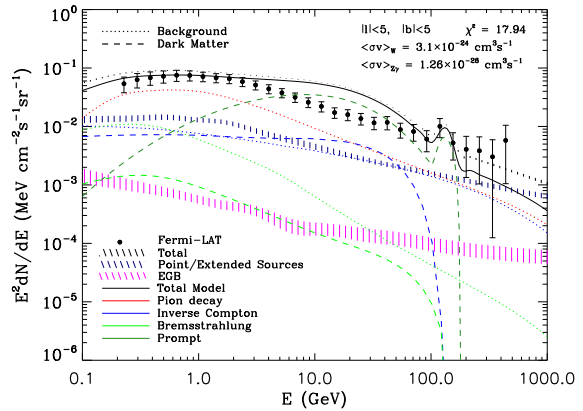}
\caption{Wino/Axion model of \cite{Acharya:2012dz}. $m_{\chi} = 145$ GeV  
$\langle \sigma v \rangle_{\chi \chi \longrightarrow Z\gamma} = 1.26 \times 10^{-26}$
cm$^{3}$s$^{-1}$, 
$\langle \sigma v \rangle_{\chi \chi}^{tot} = 3.2 \times 10^{-24}$ cm$^{3}$s$^{-1}$.}
\label{fig:Kane}
\end{figure}

In \cite{Acharya:2012dz} the DM is composed by Winos and Axions at about equal 
amounts in DM mass density towards the GC. Following \cite{Acharya:2012dz} we
take the DM mass density in Winos to be $49\%$ of the total in the GC and in the entire 
Galaxy. 
Using the Einasto model of eq.~\ref{eq:Einasto} we take DM mass to be $m_{\chi} = 145$
GeV and the cross-section to $Z\gamma$ line to be $1.26 \times 10^{-26}$. The 
total annihilation cross-section of the Winos is $3.2 \times 10^{-24}$ cm$^{3}$s$^{-1}$
and is dominantly to $W^{+}W^{-}$. 
Assuming a BR=0.96 for annihilation to $W^{+}W^{-}$ we derive that such a model is excluded
as we show in Fig.~\ref{fig:Kane}. 
In fact such a cross-section is $O(10)$ larger than the relevant 3$\sigma$ limit for 
that mass and channel shown in Fig.~\ref{fig:LimitsFull}.

We note that even ignoring the inverse Compton and bremsstrahlung components
and all the background contribution, \textit{the prompt component which also 
includes the line signal overshoots the total $\gamma$-ray spectrum between
10 and 40 GeV}. This result can not depend on just a different assumption 
for the DM profile or on a varying ratio in Wino to Axion mass density 
within the Galaxy since by changing any of these assumptions the $\gamma$-ray
line will decrease/increase by the same amount.    
 
\section{Conclusions}
\label{sec:Conclusions}

Inspired by recent indications for a monochromatic $\gamma$-ray signal
towards the GC, possibly connected to DM annihilations, we have derived 
limits on the continuous component that in general 
accompanies such signal. We use the region at $\mid l \mid < 5^{\circ}$,
$\mid b \mid < 5^{\circ}$ and compute 3$\sigma$ upper limits for DM annihilation 
cross-sections into the
$W^{+}W^{-}$, $b\bar{b}$, $\tau^{+}\tau^{-}$, $\mu^{+}\mu^{-}$ and $e^{+}e^{-}$ channels. 

We derive limits both from the total DM $\gamma$-ray emission, including the prompt,
the inverse Compton and the bremsstrahlung components as given in Fig.~\ref{fig:LimitsFull}, and 
only from prompt DM $\gamma$-rays (given in Fig.~\ref{fig:LimitsPromptONLY}). 
These limits do not depend on the exact normalization of the line(s) since they 
are dominated by the $\gamma$-ray data below 100 GeV, where the lines do not contribute
(see Table~\ref{tab:LineNorm}).
While our limits depend on the choice of the DM halo profile,
they can be easily rescaled to a different halo configuration. This happens 
since, apart from the $\mu^{+}\mu^{-}$ and $e^{+}e^{-}$ channels, 
the prompt $\gamma$-rays are the dominant component. 

We study the $\gamma$-ray data from other angular windows of the Galaxy and find that 
for cuspy DM profiles such as the NFW profile, the most stringent constraints come from our 
$\mid l \mid < 5^{\circ}$, $\mid b \mid < 5^{\circ}$ window while for the Einasto profile slightly 
stronger limits can come from $5^{\circ} < \mid b \mid < 10^{\circ}$
(see Fig.~\ref{fig:DMprofile}).
We have also introduced a new general parametrization for the DM profile to discuss how 
centrally concentrated the profile should be to give a flux compatible with the suggested GC line 
signal without violating bounds from other angular windows. We produced results for partial line-of-sight 
integration factors which are readily applicable to any dark matter profile, showing e.g. that a Burkert 
DM halo cannot be compatible (see Fig.~\ref{fig:Jlimits}).

Our limits on specific DM annihilation channels can be linearly combined 
and readily applied to most DM models in this mass range. We apply our limits to the model of 
\cite{Acharya:2012dz} and conclude that it is excluded given that its prompt 
component exceeds the total $\gamma$-ray flux (Fig.~\ref{fig:Kane}).

\vskip 0.2 in
\section*{Acknowledgments}  
The authors would like to thank Carmelo Evoli, Ran Lu and Gabrijela Zaharijas 
for valuable discussions. PU acknowledges partial support from the European Union FP7
ITN INVISIBLES (Marie Curie Actions, PITN-GA-2011-289442).
\vskip 0.05in


\bibliography{130GeVLine}
\bibliographystyle{apsrev}

\end{document}